\documentclass{PoS}

\title{Strangeness and charm at FAIR}

\ShortTitle{Strangeness and charm at FAIR}

\author{\speaker{Laura Tolos}\\
	FIAS. Goethe-Universit\"at Frankfurt am Main,\\ Ruth-Moufang-Str. 1, 60438 Frankfurt am Main, Germany \\ and
       \\ Theory Group, KVI, University of Groningen, \\ Zernikelaan 25, 9747 AA Groningen, The Netherlands\\
        E-mail: \email{tolos@kvi.nl}}

\author{Daniel Cabrera\\
Departamento de F\'{\i}sica Te\'orica II, Universidad Complutense,\\
28040 Madrid, Spain}
\author{Daniel Gamermann, Raquel Molina and Eulogio Oset\\
        Departamento de F\'{\i}sica Te\'orica and IFIC,
Centro Mixto Universidad de Valencia-CSIC,
Institutos de Investigaci\'on de Paterna, Aptdo. 22085, 46071 Valencia, Spain}
\author{Tetsuro Mizutani\\
Department of Physics, Virginia Polytechnic Institute and State University,\\
Blacksburg, VA 24061, USA}
\author{Angels Ramos\\
        Departament d'Estructura i Constituents de la Mat\`eria\\
Universitat de Barcelona,
Diagonal 647, 08028 Barcelona, Spain}

\abstract{We study the properties of strange and charm mesons in hot and dense matter within a self-consistent coupled-channel approach for the experimental conditions of density and temperature expected for the CBM experiment at FAIR/GSI. The in-medium solution at finite temperature accounts for Pauli blocking effects, mean-field binding on all the baryons involved, and meson self-energies. In the strange sector, 
the $\bar K$ spectral function spreads over a wide range of energies,
reflecting the melting of the $\Lambda (1405)$ resonance and the contribution
of $(\Lambda,\Sigma,\Sigma^*)N^{-1}$ components at finite temperature. In the case of charm mesons, the dynamically-generated $\Lambda_c(2593)$ and $\Sigma_c(2880)$ resonances remain close to their free-space position while acquiring a
remarkable width. As a result, the
$D$ meson spectral density shows a single pronounced peak for 
energies close to the $D$ meson free-space mass that
broadens with increasing matter density with an extended tail
particularly towards lower energies. We also discuss the implications for the  $D_{s0}(2317)$, $D_0(2400)$ and the predicted $X(3700)$ resonances at FAIR energies.
          }

\FullConference{8th Conference Quark Confinement and the Hadron Spectrum \\
		 September 1-6 2008\\
		 Mainz, Germany}

\begin{document}

\section{Introduction}

Strangeness has been a matter of study over the last years  in
connection to exotic atoms \cite{Friedman:2007zz} as well as heavy-ion collisions \cite{Fuchs:2005zg}. 
The analysis of the $\bar K$ interaction in nuclei has revealed some interesting features. Phenomenology of kaonic atoms shows that the $\bar K$ feels an attractive potential at low densities. This attraction results from the modified $s$-wave $\Lambda(1405)$ resonance in the medium due to Pauli blocking effects \cite{Koch} together with the self-consistent consideration of the $\bar K$ self-energy \cite{lutz} and the inclusion of self-energies of the mesons and baryons in the intermediate states \cite{Ramos:1999ku}. Attraction of the order of -50 MeV at normal nuclear matter density $\rho_0=0.17 \,{\rm fm^{-3}}$ is obtained by different approaches, such as unitarizated chiral theories in coupled-channels \cite{Ramos:1999ku}. Higher-partial waves have been studied \cite{laura,Tolos:2006ny,Lutz:2007bh,Tolos:2008di} and shown to be relevant for heavy-ion collisions. At present, relativistic heavy-ion experiments at beam energies below 2AGeV \cite{Fuchs:2005zg} have been testing strange mesons not only in a dense but also in a hot  medium. 

 In this context, the charm degree of freedom has become a recent topic of analysis. The future CBM (Compressed Baryon Matter) experiment of the FAIR (Facility of Antiproton and Ion Research) project at GSI \cite{FAIR} will investigate, among others, the  modification of the properties of open and hidden charm mesons in a hot dense baryonic environment.  The in-medium modification of the $D (\bar D)$ mesons may explain, for example, the $J/\Psi$ suppression, on the basis of a mass reduction of $D (\bar D)$ in the nuclear medium. However, a self-consistent coupled-channel calculation is needed due to the strong coupling among the
$DN$ and other meson-baryon channels \cite{TOL04,LUT06,TOL07}, which induces the appearance of dynamically-generated resonances.

In this paper we study the properties of strange ($\bar K$) and charm ($D$) mesons in a hot nuclear medium pursuing a self-consistent coupled-channel procedure. Moreover, we analyze the effect of the self-energy of $D$ mesons on dynamically-generated charm and hidden charm scalar resonances.

\section{Strange and charm mesons in hot and dense matter}

\begin{figure}[htb]
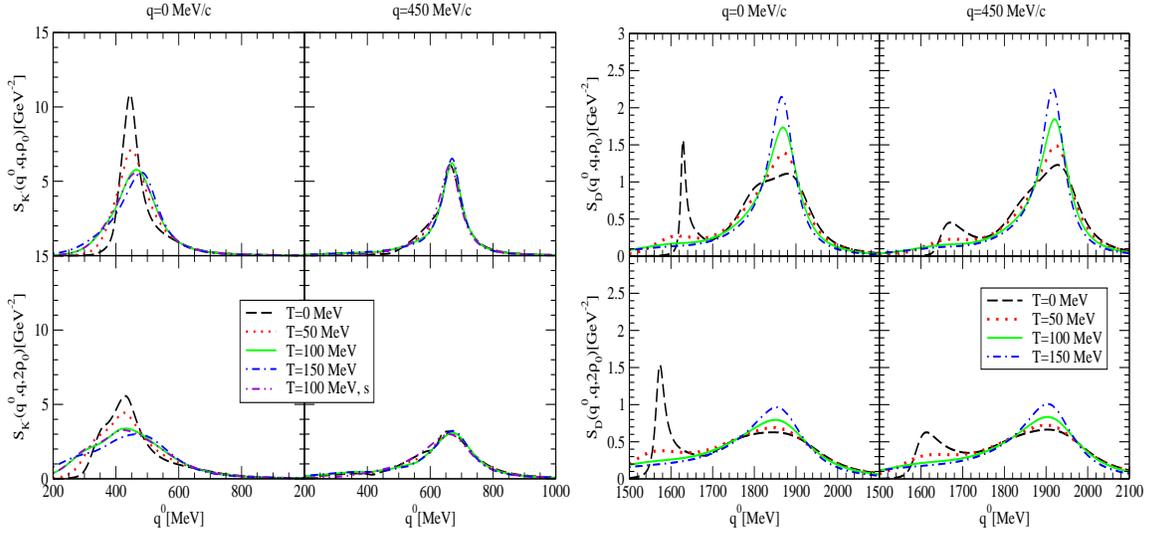

\begin{center}
\includegraphics[height=7 cm, width=7.5 cm]{spectral_tot_s+p_kbarn.eps}
\hfill
\includegraphics[height=7 cm, width=7.5 cm]{tolos_fig4.eps}
\caption{Spectral function for $\bar K$ (left) and $D$ (right) mesons.} \label{fig1}
\end{center}
\end{figure}

The self-energy and, hence, the spectral function at finite temperature for strange ($\bar K$) and charm ($D$) mesons is obtained using a self-consistent coupled-channel procedure. 

For $\bar K$, a chiral unitary approach in coupled channels is performed, which incorporates the $s$- and $p$-waves of the kaon-nucleon interaction. The $s$-wave
amplitude is obtained from the Weinberg-Tomozawa term of the $SU(3)$ chiral lagrangian.
Unitarization in coupled channels is imposed by solving the Bethe-Salpeter
equation with on-shell amplitudes. The model generates dynamically the $\Lambda
(1405)$ resonance in the $I=0$ channel. The in-medium solution at finite temperature of the $s$-wave
amplitude accounts for Pauli-blocking effects, mean-field binding on the nucleons and
hyperons via a temperature-dependent $\sigma-\omega$ model, and the dressing of
the pion and antikaon through their corresponding self-energies. This requires a
self-consistent evaluation of the $\bar K$ self-energies. The $p$-wave
self-energy is obtained via the hyperon-hole ($YN^{-1}$) excitations (see Ref.~\cite{Tolos:2008di}).

In the case of the $D$ meson, the multichannel Bethe-Salpeter equation is solved taking, as bare interaction, a type of broken SU(4) $s$-wave Tomozawa-Weinberg  interaction supplemented by an attractive isoscalar-scalar term and using a cutoff regularization scheme. This cutoff is fixed by reproducing the position and the width of the $I=0$ $\Lambda_c(2593)$ resonance. As a result, a new resonance in $I=1$ channel $\Sigma_c(2880)$ is generated \cite{LUT06}. The in-medium solution at finite temperature 
incorporates, as well, Pauli blocking effects, baryon mean-field bindings and $\pi$ and $D$ meson self-energies in a self-consistent manner (see Ref.~\cite{TOL07}). The $p$-wave self-energy is also obtained via the corresponding $Y_cN^{-1}$ excitations \cite{Molina:2008nh}.

For both cases, the self-energy is obtained self-consistently summing the transition amplitude $T$ for the different isospins over the nucleon Fermi distribution at a given temperature, $n(\vec{q},T)$, as 
\begin{eqnarray}
\Pi(q_0,{\vec q},T)= \int \frac{d^3p}{(2\pi)^3}\, n(\vec{p},T) \,
[{T}^{(I=0)} (P_0,\vec{P},T) +
3{T}^{(I=1)} (P_0,\vec{P},T)]\ , \label{eq:selfd}
\end{eqnarray}
where $P_0=q_0+E_N(\vec{p},T)$ and $\vec{P}=\vec{q}+\vec{p}$ are
the total energy and momentum of the meson-nucleon pair in the nuclear
matter rest frame, and ($q_0$,$\vec{q}\,$) and ($E_N$,$\vec{p}$\,) stand  for
the energy and momentum of the meson and nucleon, respectively, also in this
frame. The spectral function then reads
\begin{equation}
S(q_0,{\vec q}, T)= -\frac{1}{\pi}\frac{{\rm Im}\, \Pi(q_0,\vec{q},T)}{\mid
q_0^2-\vec{q}\,^2-m^2- \Pi(q_0,\vec{q},T) \mid^2 } \ .
\label{eq:spec}
\end{equation}

On the l.h.s. of  Fig.~\ref{fig1}, the $\bar{K}$ spectral function is depicted. It  shows a strong mixing between the quasi-particle peak and the $\Lambda(1405)N^{-1}$ and  $YN^{-1}$
excitations. The effect of the $p$-wave $YN^{-1}$ subthreshold excitations is
repulsive for the $\bar K$ potential, compensating in part the attraction
from the $s$-wave ${\bar K} N$ interaction.  Temperature softens the
$p$-wave contributions to the spectral function at the quasi-particle energy. Moreover, together with the $s$-wave mechanisms, the $p$-wave self-energy
provides a low-energy tail which spreads the spectral function considerably, due
to the smearing of the nucleonic Fermi surface. Density dilutes the spectral function further.

\begin{figure}[htb]
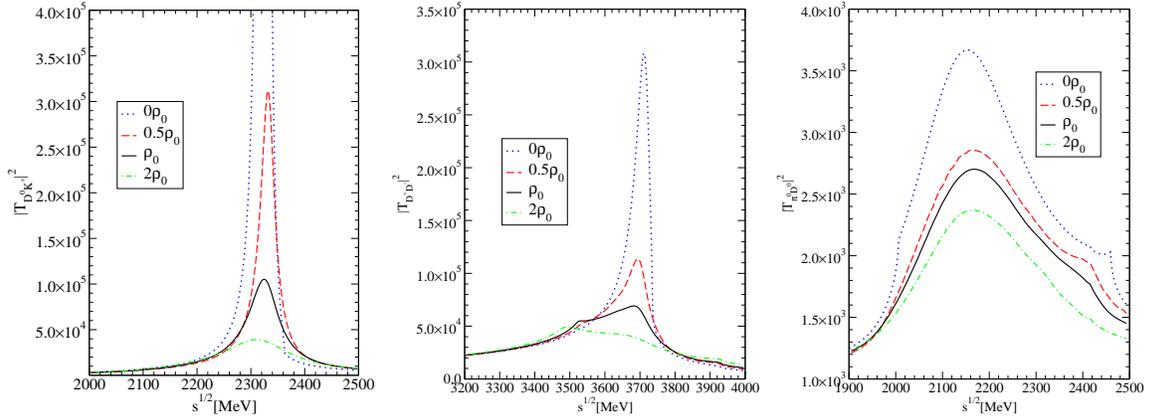

\begin{center}
\includegraphics[height=5.5 cm, width=4.9 cm]{ds02317.eps}
\hfill
\includegraphics[height=5.5 cm, width=4.9 cm]{x37.eps}
\hfill
\includegraphics[height=5.5 cm, width=4.9 cm]{d2400.eps}
\caption{$D_{s0}(2317)$ (left), $X(3700)$ (middle) and $D_0(2400)$ (right) resonances in dense matter.} \label{fig2}
\end{center}
\end{figure}

The evolution with temperature and density of the $D$ meson spectral function is shown on the r.h.s of  Fig.~\ref{fig1}. We observe the dilution of the $ \Lambda_c(2593) N^{-1}$
and $\Sigma_c(2880) N^{-1}$  structures  with increasing
temperature, while the quasiparticle peak gets closer to its free
value becoming narrower. Therefore, the distribution of 
the spectral function concentrates around the quasiparticle energy, although 
maintaining the overall strength in its lower energy part. As density
increases, the quasiparticle peak broadens and the low-energy strength,
associated to the $\Lambda_c(2593) N^{-1}$ components and related to $Y_c \pi
N^{-1}$, $Y_c N N^{-2}$, \dots absorption  modes, increases.

\section{Charm and hidden charm resonances in nuclear matter}

The renormalization of the properties of mesons in nuclear matter as, for example, the $D$ meson, can have important consequences on the medium modification of scalar mesons which are dynamically generated in the 
charm sector. Concretely, we shall study the medium modification of the
$D_{s0}(2317)$ and $D_0(2400)$. In
addition, we also study the medium changes of a hidden charm scalar
meson predicted in \cite{Gamermann:2007mu}, the $X(3700)$, which might have been observed by Belle
\cite{Abe:2007sy}.

Those resonances are generated dynamically by solving the coupled-channel Bethe-Salpeter equation for two pseudoscalars. The kernel is derivated from an 
 extrapolation to $SU(4)$ of the $SU(3)$ chiral Lagrangian used
 to generate the scalar mesons $\sigma(600)$, $f_0(980)$, $a_0(980)$ and 
 $\kappa(900)$ in the light sector, but with the $SU(4)$ symmetry strongly 
 broken, mostly due to the explicit consideration of the masses of the vector 
 mesons exchanged between pseudoscalars \cite{Gamermann:2006nm}.

The $D_{s0}(2317)$ mainly couples to $DK$ system, while the $D_0(2400)$ to $D\pi$.  On the other hand, the  hidden charm state $X(3700)$ couples most strongly to 
$D\bar{D}$. Therefore, any change in the $D$ meson properties in nuclear matter will have an important effect on those  resonances. In Fig.~\ref{fig2}, we obtain that in the case of the $D_{s0}(2317)$ and 
$X(3700)$ resonances, which have a  zero and small width,
respectively,
the medium effects  lead to widths of the order of 100 or 200
MeV at normal nuclear matter density, respectively. As for the $D_0(2400)$, we observe  an 
 extra widening from the already large width of the resonance in free space. However,
  the large original width makes the medium effects comparatively much 
  weaker than for the other two resonances \cite{Molina:2008nh}.

The study of the width of those resonances and the medium
reactions contributing to it provides information on the features of the
resonances and the self-energy of the $D$ meson in a nuclear medium. We suggest to look at transparency ratios to investigate those in-medium widths. 

 In conclusion,
the experimental analysis of those properties is a
valuable test of the dynamics of the $D$ meson interaction with nucleons and
nuclei, and the nature of the charm and hidden charm scalar resonances,
 all of 
them topics which are subject of much debate at present. The results obtained 
here should stimulate experimental work in hadron facilities, in particular at 
FAIR \cite{FAIR}, where the investigation of charm physics is one of the priorities.

\end{document}